\documentstyle[12pt]{article}

\newtheorem{statement}{Statement}

\begin{document}

\thispagestyle{empty}

\begin{center}
               RUSSIAN GRAVITATIONAL SOCIETY\\
               INSTITUTE OF METROLOGICAL SERVICE \\
               CENTER OF GRAVITATION AND FUNDAMENTAL METROLOGY\\

\end{center}
\vskip 15mm
\begin{flushright}
                                         RGS-VNIIMS-011/97
                                         \\ gr-qc/9712088

 \end{flushright}
\vskip 5mm

\begin{center}
{\large\bf Mixed Boundary Problem for the Traversable Wormhole Models}

\vskip 5mm
{\bf
M.Yu. Konstantinov }\\
\vskip 5mm
     {\em VNIIMS, 3-1 M. Ulyanovoy str., Moscow, 117313, Russia}\\
     e-mail:   konst@rgs.phys.msu.su  \\
\end{center}
\vskip 5mm

\begin{abstract}
The conditions of the traversable wormhole joining with the exterior
space-time are considered in details and the mixed boundary problem for the
Einstein equations is formulated. It is shown that, in opposite to some
declarations, the conditions of the wormhole joining with the exterior
space-time have non-dynamical nature and can not be defined by the physical
processes. The role of these conditions in the formation of the causal
structure of space-time is analyzed. It is shown that the causal structure
of the wormhole-type space-time models is independent from both the interior
and exterior energy-momentum tensors. This statement is illustrated in the
particular case of the spherical wormhole joining with flat exterior
space-time. The same conditions, which define the wormhole joining with the
exterior space-time, provide the absence of paradoxes in the models with
causality violation. It is pointed out, that the nature and physical
interpretation of the conditions of wormhole joining with the exterior
space-time and induced boundary conditions for the field variables is one of
the fundamental problems, which arise in the models with causality violation.
\end{abstract}

\vskip 40mm

\centerline{Moscow 1997}
\pagebreak

%%%%%%%%%%%%%%%%%%%%%%%%%%%%%%%%%%%%%%%%%%%%%%%%%%%%%%%%%%%%%%%%%%

\section{Introduction}

The investigation of space-time models with causality violation creates much
interest in the last years. This interest was stimulated by the series of
declarations about possibility of the creation of closed time-like curves
(CTCs) in the process of the dynamical evolution of some space-like
hypersurface \cite{morris,idn,frolnov,gott}. The statements about
unavoidable transformation of the traversable wormhole into the time machine
were made also \cite{frolnov}, so that classically any given wormhole may be
''absurdly easy turned into a time machine'' \cite{visser97}. Since the
existence of the closed time-like curves is associated usually with numerous
paradoxes \cite{deutsch,krama,krasnikov}, the part of the following papers
were devoted to finding some physical laws or new principles which may
forbid the creation of CTCs \cite{kim,frolov,hawking,visser,gppt,tanaka},
while another part of papers were devoted to the discussion of the so-called
''self-consistency conditions''~\cite{idn,fridman,carlini}, which must be
added to the usual Cauchy conditions to avoid paradoxes of time travel.

It is clear, that in general case space-time models may contain CTCs. The
most known examples of such models are G\"odel universe and Taub-NUT
space-time \cite{hawking-ellis}. Nevertheless the statements about
possibility of the creation of closed time-like curves (CTCs) in the process
of the dynamical evolution of some space-like hypersurface and unavoidable
transformation of the traversable wormhole into the time machine \cite
{morris,idn,frolnov} require more precise definition. Namely, the terms
''creation'', ''dynamical evolution'' and ''transformation'' (in particular,
the term ''unavoidable transformation'') are usually applied to the models
which may be obtained in a result of development of some initial
configuration or, equivalently, to the solutions of the appropriate Cauchy
problem. Application of these terms to space-time models with CTCs and time
machine contradict to the well known theorems about global hyperbolicity
\cite{hawking-ellis}, which have pure topological nature and state that
globally hyperbolic space-time has topology of the direct product $T\times
M^3$, where $T$ is a global time axes and $M^3$ - some 3-manifold. It is
clear, that globally hyperbolic space-time, which may be considered as a
global solution of appropriate Cauchy problem, can not contain CTC \cite
{hawking-ellis}. The same conclusion may be derived from the dynamical
(3+1)-formalism in general relativity \cite{mtu,york,fisher} which excludes
the existence of CTCs in the hyprbolic region. If space-time is not globally
hyperbolic then CTCs may exist in non hyperbolic regions but CTCs in such
models can not be considered as the result of transformation, dynamical
evolution or development of some initial configuration, because any non
globally hyperbolic space-time solves some boundary or mixed boundary
problem which can not be reduced to the Cauchy problem. Unfortunately, no
attention was paid on such contradiction, because all existing examples of
traversable wormholes, including the models with causality violation, were
obtained ''by hand'' without solution or analysis of the appropriate
boundary problem for the Einstein equations.

In the recent papers \cite{myuk92,myuk95} it was shown that in opposite to
the statements of \cite{morris,idn} the motion of the wormhole' mouths and
twin paradox do not lead to the transformation of the traversable Lorentzian
wormhole into the time machine. It was mentioned also, that the causality
violation and the existence of CTCs depend on the boundary conditions which
are defined by the manifold structure \cite{myuk92,myuk95} and provide the
absence of any paradoxes in the presence of CTCs. In more details the nature
of self-consistency conditions was considered in \cite{konst97}, where it
was shown that they are induced by the manifold structure and can not be
considered nor as additional conditions nor as the consequence of the
principle of minimal action as it states in \cite{carlini}. It was stressed
out also \cite{myuk92,myuk95,konst97}, that in opposite to some
declarations, the space-time models with causality violation are the subject
of the mixed boundary problem for the Einstein equations. Nevertheless in
general form this boundary problem was not considered or formulated
explicitly in these papers.

In the present paper we consider the boundary conditions which arise due to
the wormhole joining with the exterior space-time and discuss the connection
of the corresponding mixed boundary problem with the causal properties of
space-time. In the next section the general topological structure of the
traversable wormhole models is briefly described. The conditions of the
wormhole joining with the exterior space-time are considered in section 3
and the mixed boundary problem for such models is considered in section 4.
The connection between the conditions of the wormhole joining with the
exterior space and causal properties of space-time is discussed in section
5. Simple particular model is considered in section 6. The last section
contains short summary and discussion.

\section{The topological structure of the traversable wormhole models}

The general space-time model with traversable wormhole may be considered as
the result of attaching of the interior space-time with the topology $%
M_{int}^4=T_{int}\times M_{int}^3$ and exterior space-time $M_{ext}^4=
T_{ext}\times M_{ext}^3$. Here $T_{int}$ and $T_{ext}$ are the interior and
exterior time-like axes, $M_{int}^3$ and $M_{ext}^3$ are the interior and
exterior spaces, which satisfy to the following conditions:

\begin{itemize}
\item  $M_{int}^3$ is a connected compact orientable 3-manifold whose
boundary $\partial M_{int}^3$ consists of two compact orientable 2-manifolds
$M_l^2$ and $M_r^2$ with empty intersection ($M_l^2\cap M_r^2=\emptyset $);

\item  the intersection $M_{int}^4\cap M_{ext}^4=T_{int}\times (M_l^2\cup
M_r^2)=T_{ext}\times (M_l^2\cup M_r^2)$;

\item  both $T_{int}\times M_l^2=T_{ext}\times M_l^2$ and $T_{int}\times
M_r^2=T_{ext}\times M_r^2$ belong to the same connected component of $%
M_{ext}^4$.
\end{itemize}

The interior 3-space $M_{int}^3$ is called a handle of the wormhole and the
manifolds $M_l^2$ and $M_r^2$ are often called as the ''left'' and ''right''
mouths of the wormhole. According to well-known theorems~\cite{milnor}, $%
M_l^2$ and $M_r^2$ may be arbitrary 2-manifolds with genuses $p_l$ and $p_r$%
, i.e. they are diffeomorphic to the 2-sphere $S^2$ with $p_l$ and $p_r$
handles respectively, and the manifold $M_{int}^3$ is an interpolating
manifold.

In the simplest models \cite
{morris,idn,frolnov,myuk92,myuk95,konst97,diaz,thorne} the interior space of
the wormhole has topological structure of the direct product $%
M_{int}^3=I\times M^2$ of interval $I=(-L_1,L_2)$ and a compact orientable
2-dimensional manifold $M^2$, so that the ''left'' and ''right '' mouths of
the wormhole have the same topology $M_l^2=M_r^2=M^2$. The sum $L=L_1+L_2$
is called a length (to be exact, a coordinate length) of the wormhole's
handle.

The attaching of the interior and exterior space-times is made by the
following standard manner. Let $t\in T_{ext}$ and $\tau \in T_{int}$ are the
exterior and interior time coordinates. Then, for fixed $\tau $ one (left)
wormhole mouth is attached to the space-like section $M_{ext1}^3=(t_1(\tau
),M_{ext}^3)$ while another (right) mouth is attached to the section $%
M_{ext2}^3=(t_2(\tau ),M_{ext}^3)$. The attaching of the wormhole mouths to
the exterior spaces $M_{ext1}^3$ and $M_{ext2}^3$ are made as follows: to
attach the wormhole along its left mouth $(-L_1,M_l^2)$ to the exterior
space $M_{ext1}^3$ it is necessary to remove the tubular neighborhood of $%
M_l^2\in M_{ext1}^3$ from the exterior space $M_{ext1}^3$ and join the
boundary $M_l^2$ of the rest of $M_{ext1}^3$ with the left wormhole mouth $%
(-L_1,M_l^2)$. The same procedure is used for the joining of the right
wormhole mouth $(L_2,M_l^2)$ with the exterior space $M_{ext2}^3$. In the
result of such joining the wormhole connects the points of exterior
space-like hypersurface $(t_1,M_{ext1}^3)$ with the points of exterior
space-like hypersurface $(t_2,M_{ext2}^3)$, where $t_1\neq t_2$ in general
case.

\section{Boundary conditions for the traversable wormhole models}

To discuss the causal structure of the wormhole-type models it is more
suitable to use local coordinate consideration. Consider for this purpose
the curve $l\in M_{int}^3$ which pass through the wormhole handle $M_{int}^3$
and connects two points $p_l\in M_l^2$ and $p_r\in M_r^2$ of the left and
right mouths of the wormhole. Consider the tubular neighborhood $%
U_{int}\subset M_{int}^3$ of the curve $l$ in $M_{int}^3$, which has
topology of the direct product $U_{int}=l\times D^2$ of the line $l$ and an
open disk $D^2\subset R^2$. Let $\{\tau ,\xi ^1,\xi ^2,\xi ^3\}$ are the
local coordinates in $T_{int}\times U_{int}$, such that $-\infty <\tau
<\infty $ is an interior time-like coordinate, $\xi ^1$ is the coordinate
along the line $l$, $-(\sigma _1+L_1)<\xi ^1<L_2+\sigma _2$, where $\sigma
_1 $, $\sigma _2$, $L_1$ and $L_2$ are some positive constants, values $\xi
^1=-L_1$ and $\xi ^1=L_2$ correspond to the left and right mouths of the
wormhole respectively, the regions $-(\sigma _1+L_1)<\xi ^1<L_1$ and $L<\xi
^1<L+\sigma _2$ correspond to the wormhole intersection with the exterior
space-time, $\xi ^2$ and $\xi ^3$ are the coordinates in the disk $D^2$.

In general form the metric of space-time in the wormhole interior may be
written in the coordinates $\{\tau ,\xi ^1,\xi ^2,\xi ^3\}$ as follows
\begin{equation}
\label{intmetric}ds_{int}^2=a^2(\tau ,\xi )d\tau ^2-2b_i(\tau ,\xi )d\tau
d\xi ^i-\widetilde{\gamma }_{ij}(\tau ,\xi )d\xi ^id\xi _j
\end{equation}
where $\xi =\{\xi ^1,\xi ^2,\xi ^3\}$, $\widetilde{\gamma }_{ij}(\tau ,\xi )$
denotes the metric of the interior 3-space $\tau =const$, $a^2(\tau ,\xi )>0$
because of the supposition about the wormhole traversability and $b_i(\tau
,\xi )\neq 0$ in general case. The limiting values of the function $a^2(\tau
,\xi )$, 3-vector $b_i(\tau ,\xi )$ and 3-tensor $\widetilde{\gamma }%
_{ij}(\tau ,\xi )$ at $\xi ^1\rightarrow -L_1,\,L_2$ are defined by the
wormhole joining with the exterior space-time as it describes below.

Without loss of generality it may be supposed that in the exterior
space-time both wormhole mouths are covered by the same map $%
\{t,x^1,x^2,x^3\}$, where $-\infty <t<\infty $ is an exterior time and $%
\{x^1,x^2,x^3\}$ are the coordinates on the exterior space section $t=const$%
. For simplicity it will be supposed additionally, that the coordinates $%
\{t,x^1,x^2,x^3\}$ of the exterior space-time are synchronous, so, the
exterior space-time metrics may be written as
\begin{equation}
\label{extmetr}ds_{ext}^2=dt^2-\gamma _{ij}dx^idx^j
\end{equation}
where $\gamma _{ij}$ is a positive definite metric of the exterior 3-space $%
t=const$.

Now it is necessary to define the wormhole joining with the exterior
space-time. Without loss of generality it may be supposed, that the exterior
coordinates are comoving to the left mouth of the wormhole, so, for $%
-(\sigma _1+L_1)<\xi ^1<-L_1$ we have
\begin{equation}
\label{extleft1}t_{left}=\tau ,
\end{equation}
\begin{equation}
\label{extleft2}x_{left}^i=x^i(\xi ^1,\xi ^2,\xi ^3)
\end{equation}
while for the right mouth, i.e. for $L_2<\xi ^1<L_2+\sigma _2$, we have in
general
\begin{equation}
\label{extright1}t_{right}=t_r(\tau ),
\end{equation}
\begin{equation}
\label{extright2}x_{right}^i=x^i(\tau ,\xi ^1,\xi ^2,\xi ^3)
\end{equation}
These equations define the joining of the interior space-time of the
wormhole with the exterior space-time. Equations (\ref{extleft1}) and (\ref
{extright1}) show that wormhole connects the space section $t=\tau $ of the
exterior space-time with the space section $t=t_r(\tau )$ which, in general,
does not coincide with space section $t=\tau $.

It is necessary to note, that in several wormhole models instead of the
condition (\ref{extright1}) more general condition $t_{right}=t_r(\tau ,\xi
^i)$ is used \cite{morris,idn}. In this case different points of the right
wormhole's mouth (i.e. the points with coordinates $\{L_2,\xi ^2,\xi ^3\}$
are placed on different space sections of the exterior space-time. It is
easy to see, that such generalization has no influence on the causal
structure of space-time and by this reason it will not considered here.

Using the joining equations (\ref{extleft1})-(\ref{extright2}) near the left
wormhole mouth (i.e. for $-(\sigma _1+L_1)<\xi ^1<-L_1$) the exterior
space-time metric (\ref{extmetr}) may be rewritten in the interior
coordinates $\left\{ \tau ,\xi ^1,\xi ^2,\xi ^3\right\} $ as
\begin{equation}
\label{extmetrleft}ds_{ext}^2=d\tau ^2-\gamma _{ij}^{\prime }d\xi ^id\xi ^j
\end{equation}
where
\begin{equation}
\label{spaceleft}\gamma _{ij}^{\prime }=\gamma _{kl}x_i^kx_j^l,\qquad
x_i^k=\partial x^k/\partial \xi ^i,
\end{equation}
while near the right mouth ($L_2<\xi ^1<L_2+\sigma _2$) it takes the form
\begin{equation}
\label{rightmetr}ds_{ext}^2=\alpha ^2(1-v_lv^l)d\tau ^2-2\beta _id\tau d\xi
^i-\gamma _{ij}^{\prime \prime }d\xi ^id\xi ^j
\end{equation}
where
\begin{equation}
\label{boundright}\alpha =dt_r/d\tau ,\quad v_l=\frac 1\alpha \gamma _{kl}%
\frac{\partial x^k}{\partial \tau },\quad \beta _i=\gamma _{kl}\frac{%
\partial x^k}{\partial \tau }\frac{\partial x^l}{\partial \xi ^i},
\end{equation}
\begin{equation}
\label{spaceright}\gamma _{ij}^{\prime \prime }=\gamma _{kl}x_i^kx_j^l,
\end{equation}
$v_i$ is a vector of the ''3-velocity'' of the right mouth in the
coordinates $\{t,x^i\}$ and $x_i^k=\partial x^k/\partial \xi ^i$ with $%
x^k=x^k($ $\tau ,\xi ^1,\xi ^2,\xi ^3)$. The traversability condition gives
the following restrictions on the functions $t_r(\tau )$ and $x^i($ $\tau
,\xi )$:
$$
\alpha ^2(1-v_lv^l)>0,
$$
and hence
\begin{equation}
\label{trans1}\alpha ^2>\varepsilon >0,
\end{equation}
where $\varepsilon =const>0$, i.e. $t_r(\tau )$ must be a monotonous
function on $\tau $ without stationary points, and
\begin{equation}
\label{trans2}v_lv^l<1.
\end{equation}

Equations (\ref{extmetrleft})-(\ref{spaceright}) with constraints (\ref
{trans1})-(\ref{trans2}) define the boundary conditions for the interior
wormhole metrics (\ref{intmetric}). Namely, the comparison of (\ref
{intmetric}) with its limiting forms (\ref{extmetrleft}) and (\ref{rightmetr}%
) gives
\begin{equation}
\label{asympt1}a^2(\tau ,\xi )=%
\cases{1 & for $-(\sigma _1+L_1)<\xi ^1<-L_1$; \cr
\alpha ^2\cdot (1-v_iv^i) & for $L_2<\xi ^1<L_2+\sigma _2$, \cr
}
\end{equation}

\begin{equation}
\label{asympt2}\beta _i=%
\cases{
0 & for $-(\sigma _1+L_1)<\xi ^1<-L_1$ \cr
\gamma _{kl}\frac{\partial x^k}{\partial \tau }\frac{\partial x^l}{\partial
\xi ^i} & for $L_2<\xi ^1<L_2+\sigma _2$, \cr
}
\end{equation}
and
\begin{equation}
\label{asympt3}\widetilde{\gamma }_{ij}=%
\cases{
\gamma _{ij}^{\prime }, & for $-(\sigma _1+L_1)<\xi ^1<-L_1$ \cr
\gamma _{ij}^{\prime \prime } & for $ L_2<\xi ^1<L_2+\sigma _2$. \cr
}
\end{equation}
where $\gamma _{kl}$ are the metric tensor of the exterior 3-space, $%
x^k=x^k( $ $\xi ^i)$ near the left mouth ($\xi ^1\rightarrow -(\sigma
_1+L_1) $) and $x^k=x^k($ $\tau ,\xi ^i)$ near the right mouth ($\xi
^1\rightarrow L_2$), $\gamma _{ij}^{\prime }$ and $\gamma _{ij}^{^{\prime
\prime }}$ are defined by the equations (\ref{spaceleft}) and (\ref
{spaceright}).

\section{Mixed boundary problem for the wormhole-type models}

Now we may formulate the mixed boundary problem for the traversable wormhole
models.

It is clear, that to construct any space-time model it is necessary to have
the following data: structure of manifold, field equations, initial and/or
boundary conditions.

{\bf 1. }{\it The structure of manifolds. }According to the above assumption
space-time consist of the internal and external parts, which have topologies
$T_{int}\times M_{int}^3$ and $T_{ext}\times M_{ext}^3$ respectively, where $%
T_{int}$ and $T_{ext}$ are the interior and exterior times and $M_{int}^3$
and $M_{ext}^3$ are the interior and exterior spaces. Up to the definition
of $M_{int}^3$ and $M_{ext}^3$, the structure of space-time, i.e. the
joining of interior and exterior parts, is given by the equations (\ref
{extleft1})-(\ref{extright2}) which must satisfy to the constraints (\ref
{trans1}) and (\ref{trans2}).

{\bf 2. }{\it Field equations.} For simplicity, it will supposed that the
metric of space-time must satisfy to the standard Einstein equations
\begin{equation}
\label{E00}G_0^0=\kappa T_0^0,\qquad G_i^0=\kappa T_i^0
\end{equation}
and
\begin{equation}
\label{Eij}G_j^i=\kappa T_j^i
\end{equation}
where $G_\beta ^\alpha $ is Einstein tensor, $\kappa $ is Einstein
gravitational constant and $T_\beta ^\alpha $ is the energy-momentum tensor
of matter and non-gravitational fields. Equations (\ref{E00}) are the
constraint equations and equations (\ref{Eij}) are dynamical. These
equations must be supplemented by the matter and non-gravitational fields
equations, which will not considered here.

{\bf 3.} {\it Initial and boundary conditions.} According to the assumptions
about the interior and exterior space-time structures and the traversability
of the wormhole, the interior metric must have the form (\ref{intmetric})
with $a^2(\tau ,\xi )>\varepsilon >0$ and it is supposed for simplicity that
the exterior space-time metric has the form (\ref{extmetr}). Equations (\ref
{extleft1})-(\ref{extright2}), which define the topology of space-time,
induce the correspondence between components of the interior and exterior
metric tensors in the intersection of interior and exterior regions of
space-time. Equations (\ref{asympt1})-(\ref{asympt3}), which define this
correspondence, must be considered as additional boundary conditions for the
components of the metric tensor.

Indeed, in the regions $-(\sigma _1+L_1)<\xi ^1<-L_1$ and $L_2<\xi
^1<L_2+\sigma _2$, where interior space-time intersects with exterior one,
both the equations (\ref{extleft1})-(\ref{extright2}) and induced equations (%
\ref{asympt1})-(\ref{asympt3}) have the form of coordinate transformation
and have no effect on the energy-momentum tensor. Therefore, in opposite to
wide spread opinion~\cite
{morris,idn,frolnov,kim,frolov,hawking,fridman,carlini,diaz}, these equation
are independent from the field equations and hence, have nondynamical nature.

At last, usual initial conditions for the Einstein equations must be given,
namely: the components of the interior metric tensor and their first partial
derivatives with respect to the interior time coordinate $\tau $, i.e.
\begin{equation}
\label{initint}a^2(\tau _0,\xi ),\;\beta _i(\tau _0,\xi ),\;\widetilde{%
\gamma }_{ij}(\tau _0,\xi ),\;\partial a^2(\tau _0,\xi )/\partial \tau
,\;\beta _{i,\tau }(\tau _0,\xi ),\;\widetilde{\gamma }_{ij,\tau }(\tau
_0,\xi ),
\end{equation}
with%
$$
a^2(\tau _0,\xi )>0,
$$
and the components of the exterior metric and their first partial
derivatives with respect to the exterior time coordinate $t$, i.e.
\begin{equation}
\label{initext}\gamma _{ij}(t_0,x),\quad \gamma _{ij,t}(t_0,x).
\end{equation}
These quantities must satisfy to the boundary conditions (\ref{asympt1})-(%
\ref{asympt3}) and the corresponding conditions for derivatives, which near
the left mouth ($-(\sigma _1+L_1)<\xi ^1<-L_1$) take the form
\begin{equation}
\label{inleft1}\frac{\partial a^2(\tau _0,\xi )}{\partial \tau }=0,\qquad
\frac{\partial \beta _i(\tau _0,\xi )}{\partial \tau }=0,
\end{equation}
and
\begin{equation}
\label{inleft2}\widetilde{\gamma }_{ij,\tau }(\tau _0,\xi )=\gamma
_{kl,t}(t_0,x)x_i^kx_j^l
\end{equation}
while near the right mouth we have
\begin{equation}
\label{inright1}\frac{\partial a^2(\tau _0,\xi )}{\partial \tau }=2\alpha
\alpha _\tau \cdot (1-v_iv^i)-2\alpha ^2v_{i,\tau }v^i
\end{equation}
\begin{equation}
\label{inright2}\beta _{i,\tau }(\tau _0,\xi )=\alpha \gamma _{kl,t}\frac{%
\partial x^k}{\partial \tau }\frac{\partial x^l}{\partial \xi ^i}+\gamma
_{kl}\frac{\partial ^2x^k}{\partial \tau ^2}\frac{\partial x^l}{\partial \xi
^i}+\gamma _{kl}\frac{\partial x^k}{\partial \tau }\frac{\partial ^2x^l}{%
\partial \xi ^i\partial \tau }
\end{equation}
and
\begin{equation}
\label{inright3}\widetilde{\gamma }_{ij,\tau }=\alpha \gamma _{kl,t}\frac{%
\partial x^k}{\partial \xi ^i}\frac{\partial x^l}{\partial \xi ^j}+2\gamma
_{kl}\frac{\partial ^2x^k}{\partial \xi ^i\partial \tau }\frac{\partial x^l}{%
\partial \xi ^j}.
\end{equation}
It is obviously, that these quantities must satisfy also to the constraint
equations (\ref{E00}).

The contents of this section may be summarized as follows:

\begin{statement}
Any space-time model with the traversable wormhole whose interior and
exterior metrics have the forms (\ref{intmetric}) and (\ref{extmetr})
respectively, is the subject of the mixed boundary problem for the Einstein
equation (\ref{E00})-(\ref{Eij}) which is formed by (i) the manifold
structure equations (\ref{extleft1})-(\ref{extright2}), (ii) the boundary
conditions (\ref{asympt1})-(\ref{asympt3}) with the traversability
constraints (\ref{trans1})-(\ref{trans2}), (iii) interior and exterior
initial conditions (\ref{initint})-(\ref{initext}) which must satisfy to (%
\ref{asympt1})-(\ref{asympt3}) and (\ref{inleft1})-(\ref{inright3}).
\end{statement}

\section{Causality violation in the wormhole models}

The above equations make possible to obtain some estimations for the
causality violation in the traversable wormhole models. Without loss of
generality it may be supposed that the sizes of the mouths are much less
than the distance between them in the outer space (the approximation of the
thin mouths) and the $x^1$ axis connects the centers of the wormhole's
mouths. Let moreover, $t_r(\tau _1)>\tau _1$ for some $\tau _1>\tau _0$.
Consider the light signal which is sent at the moment $t_r(\tau _1)$ from
the right mouth to the left one through the wormhole and then return to the
right mouth through the outer space. It is clear, that the time delay
between the sending and receiving of the signal is equal to%
$$
\Delta t=\delta t_1+\delta t_2-\delta t_3
$$
where $\delta t_1$ and $\delta t_2$ are the times of the signal passing
through the wormhole and the exterior space respectively and $\delta t_3=$ $%
t(\tau _1)-\tau _1$. It is clear that causality violation appears if $\Delta
t\leq 0$. The estimations of the times $\delta t_1$ and $\delta t_2$ may be
obtained from the equations (\ref{intmetric}) and (\ref{extmetr}). Namely,
let%
$$
R=x^1(\tau _1,L_2),\qquad C_{ext}=\max _{0\leq x^1\leq R}\gamma _{11}(t,x),
$$
and
$$
a_0^2=\min _{-L_1\leq \xi ^1\leq L_2}a^2(\tau ,\xi ),\quad b=\max _{-L_1\leq
\xi ^1\leq L_2}\left| b_i(\tau ,\xi )\right| ,\quad N=\max _{-L_1\leq \xi
^1\leq L_2}\widetilde{\gamma }_{ij}(\tau ,\xi ),
$$
then
$$
\delta t_1\leq C_{int}L,\qquad \delta t_2\leq C_{ext}R,
$$
where $L=L_1+L_2$, and
$$
C_{int}=\frac{b+\sqrt{b^2+N}}{a_0^2},
$$
so,
$$
\Delta t\leq C_{int}L+C_{ext}R-\left| t_r(\tau _1)-\tau _1\right| .
$$
Thus, the sufficient condition for the causality violation may be written in
the form
\begin{equation}
\label{cv}\left| t(\tau _1)-\tau _1\right| \geq C_{int}L+C_{ext}R
\end{equation}
It is necessary to note that this estimation is very rough, so the causality
violation may occur even if the inequality (\ref{cv}) does not satisfied.

Analogously, if
\begin{equation}
\label{causal}\left| t_r(\tau _1)-\tau _1\right| <C_{1int}L+C_{1ext}R
\end{equation}
where%
$$
C_{1int}=\frac{b_m+\sqrt{b_m^2+N_m}}{a_1^2},\quad C_{1ext}=\min _{0\leq
x^1\leq R}\gamma _{11}(t,x),
$$
and%
$$
a_1^2=\max _{-L_1\leq \xi ^1\leq L_2}a^2(\tau ,\xi ),\quad b_m=\min
_{-L_1\leq \xi ^1\leq L_2}\left| b_i(\tau ,\xi )\right| ,\quad N_m=\min
_{-L_1\leq \xi ^1\leq L_2}\widetilde{\gamma }_{ij}(\tau ,\xi ),
$$
for all $\tau \in (-\infty ,\infty )$ then there are no CTCs in the
considered model.

The analogous estimations may be applied also to the wormhole with finite
sizes of the mouths if we suppose, that the $x^1$ axis connects two points $%
p_l$ and $p_r$ of the left and right mouths such that $R\geq
x^1(p_r)-x^1(p_l)=\min \left( x^1(p_2)-x^1(p_1)\right) $ where $p_1$ and $%
p_2 $ are the arbitrary points of the left and right mouths.

Thus, the main parameters, which define the causal structure of the
wormhole-type models with the given function $t_r(\tau )$ are the interior
''coordinate length'' $L$ of the wormhole's handle, the exterior
''coordinate distance'' $R$ between its mouths and the factors $C_{int}$ and
$C_{ext}$. It is follows from the above consideration, that parameters $L$
and $R$ are the subject of the boundary conditions for space-time models
with traversable wormhole. These parameters are independent on each other,
on the field equations and on the function $t_r(\tau )$. So, using the
appropriate choice of the parameters $L$ and $R$ (conditions (\ref{extleft2}%
) and (\ref{extright2})) both causal and non-causal space-time models with
traversable wormhole may be obtained for the same $t_r(\tau )\neq \tau $. Of
cause, for the given boundary conditions (\ref{extleft1})-(\ref{extright2})
with $t_r(\tau )\neq \tau $ the causality violation depends on the factors $%
C_{int}$ and $C_{ext}$ which are defined by the field equations. On the
other hand, if $t_r(\tau )\equiv \tau $ then causality violation is
impossible in the considered model in opposite to the statement of \cite
{frolnov} about unavoidable wormhole transformation into the time machine.

It confirm our early statements about non-dynamical nature of CTCs and the
impossibility of the dynamical wormhole transformation into the time machine
\cite{myuk92,myuk95}.

\section{Spherical wormhole in Minkowskian space-time}

To demonstrate that the causality violation has no direct connection with
some physical processes consider particular case of the traversable
spherical wormhole with immovable mouths which is joined with flat
Minkowskian exterior space-time.

The exterior flat region of such model is described by Cartesian coordinates
$\left\{ t,x,y,z\right\} $, which vary from $-\infty $ up to $\infty $, and
metric%
$$
ds^2=dt^2-dx^1-dy^2-dz^2,
$$
while the interior region is described by the coordinates $\left\{ \tau
,l,\theta ,\phi \right\} $, $-\infty <\tau <\infty $, $-(\sigma
_1+L_1)<l<L_2+\sigma _2$, and $(\theta ,\phi )$ are the polar coordinates on
2-sphere $S^2$. If the mouths of the wormhole are placed on the $x$-axis
with the centers at $x_{left}=0$ and $x_{right}=R=const$ then the joining
conditions (\ref{extleft1})-(\ref{extright2}) reads%
$$
t_{left}=\tau ,\,x_l=l\cos (\phi )\cos (\theta ),\,y_l=l\cos (\phi )\sin
(\theta ),\,z_l=l\sin (\phi )
$$
and%
$$
t_{right}=t_r(\tau ),\,x_r=R+l\cos (\phi )\cos (\theta ),\,\,y_l=l\cos (\phi
)\sin (\theta ),\,z_l=l\sin (\phi ).
$$
So, the simplest interior metric which satisfies to the boundary conditions (%
\ref{extleft1})-(\ref{extright2}) has the form
\begin{equation}
\label{sphworm}ds^2=a^2(\tau ,l)d\tau ^2-dl^2-r^2(l)(d\theta ^2+\sin
^2(\theta )d\phi ^2),
\end{equation}
with $a(\tau ,l)=1$ for $l<L_1$ and $a(\tau ,l)=dt_r/d\tau $ for $l>L_2$ and
$r(l)=l$ for $l<-L_1$ or $l>L_2$. In the particular case then $a=a(l)$ this
metric coincide with the static metric, considered in~\cite
{morris,idn,thorne}.

Direct calculation gives the following values of non-zero components of the
Einstein tensor in the wormhole interior:%
$$
G_0^0=-\frac{2rr^{\prime \prime }+r^{\prime 2}-1}{r^2}
$$
$$
G_1^1=-\frac{ar^{\prime 2}+2ra^{\prime }r^{\prime }-a}{ar^2}
$$
$$
G_2^2=G_3^3=-\left( \frac{r^{\prime \prime }}r+\frac{a^{\prime }}a\frac{%
r^{\prime }}r+\frac{a^{\prime \prime }}a\right)
$$
where the prime ($^{\prime }$) denote the partial derivative with respect to
$l$.

It is easy to see, that Einstein tensor $G_\beta ^\alpha $ and hence the
energy-momentum tensor for the interior space in this model in non-static
case ($t_{left}=\tau $, $t_{right}=t_r(\tau )\neq \tau $) have the same
structure and properties as in static case ($a(\tau ,l)=a(l)$, $%
t_{left}=t_{right}=\tau $) which was considered in~\cite{thorne}. In
particular, the matter in the wormhole interior must have the same
''exotic'' properties both in static and non-static cases. Further, the
Einstein equations impose no restrictions on the dependence of $a(\tau ,l)$
on the interior time $\tau $. Taking into consideration that $t_r(\tau )$ is
an arbitrary monotonous function ($dt_r(\tau )/d\tau $ $\neq 0$) it may
conclude that Einstein equations (and hence, the physical processes in
space-time) have no influence on the causal structure of the considered
model.

It is easy to see, that the same result may be obtained for any models with
arbitrary static exterior space-time, in particular, for the so-called
ringhole model, which was considered recently in \cite{diaz}.

\section{Conclusion}

Any space-time model in classical general relativity solves some initial
values, boundary or mixed boundary problem for the Einstein equations. Exact
formulation of such problem is very important for the physical
interpretation of the model. For the most models, which are usually
considered in the cosmological or astrophysical context, formulation of the
corresponding problem is trivial enough, because such models, as a rule,
have rather simple topology and may be considered in the framework of the
Cauchy problem with appropriate asymptotic conditions on spatial infinity.
In more complicated cases, in particular, for the topologically nontrivial
models with nontrivial causal structure, such consideration is impossible.
These models are described by means of finite or countable set of coordinate
maps \cite{hawking-ellis,milnor,kobayashi} which are joined with each other.
The conditions of the maps joining, which define the topological structure
of space-time, induce the appearance of the additional boundary conditions
(or additional constraints) for the field variables.

In this paper such conditions were considered for the space-time models with
traversable wormhole. It was shown that the conditions, which define
wormhole joining with the outer space, induce the boundary (or consistency)
conditions for the fields variables. Although these conditions in some cases
may have the form of the motion equations, they have nondinamical nature and
are independent on the fields equations. These conditions, together with
appropriate initial conditions, compose the mixed boundary problem for the
field equations. In the present paper this problem was formulated in
explicit form for the Einstein equations and its generalization on the other
fields is straightforward. The estimated conditions for the causality
violations in the traversable wormhole models show that in opposite to wide
spread opinion \cite{morris,idn,frolnov,frolov,hawking,fridman,carlini}
there is no direct connection between the causal properties in space-time
and the physical processes in it. In particular, example of the spherical
wormhole connected with exterior flat Minkowskian space-time shows that the
energy-momentum tensor in the wormhole interior may have the same properties
both in causal and noncausal cases. Moreover, if wormhole connects the
events on the same exterior space-like section, i.e. if conditions (\ref
{extleft1}) and (\ref{extright1}) have the form $t_{left}=t_{right}=\tau $,
causality violation is impossible independently on motion of the wormhole'
mouths and on the other physical processes in space-time.

Since the space-time models with traversable wormhole in general case is the
subject of the mixed boundary problem, the physical processes in such
space-time must be analyzed in the framework of the analogous problem. As it
was shown in \cite{konst97}, such consideration does not lead to any
paradoxes in the case of causality violation.

It is necessary to note that the serious problem which arise in the wormhole
models is the nature and physical interpretation of the conditions of
wormhole joining with exterior space-time (\ref{extleft1})-(\ref{extright2})
and the induced conditions for the field variables (\ref{asympt1})-(\ref
{asympt3}) in general case $t_{right}\neq t_{left}$. Really, the mixed
boundary problem for the traversable wormhole reduces to the usual Cauchy
problem only for the globally hyperbolic models without causality
violations. However, excluding the trivial case $t_{left}=t_{right}=\tau $,
the noncausal globally hyperbolic nature of the model may be determine, in
general, only after solution of the mixed boundary problem which is
formulated above. To have the full set of initial and boundary conditions
for this problem the local observer must receive information both from his
past and future. The last is very problematic, because all known physical
processes make possible to send information only from past to future, but
not reverse.

\section*{Acknowledgments}

This work was supported by the Russian Ministry of Science and the Russian
Fund of Basic Research (grant N 95-02-05785-a).

\end{document}